\documentclass[twocolumn,english,aps,prb,floatfix,amssymb,showpacs,superscriptaddress]{revtex4}
\usepackage{amsmath}
\usepackage{dsfont}
\usepackage{amssymb}
\usepackage{graphicx}
\usepackage{braket}

\ifx\pdftexversion\undefined
\usepackage[dvips]{hyperref}
\else
\usepackage{hyperref}
\fi
\hypersetup{
  colorlinks = true, linkcolor = magenta
}

\newcommand{\ssm}{\scriptscriptstyle\rm}
\newcommand{\pdag}{\phantom{\dag}}
\renewcommand{\theta}{\vartheta}
\renewcommand{\phi}{\varphi}

\begin{document}

\title{$\mathds Z_2$ slave-spin theory of a strongly correlated Chern insulator}
\begin{abstract}
We calculate the phase diagram of the topological honeycomb model in the presence of strong interactions. We concentrate on half filling and employ a ${\mathds Z}_{2}$ slave-spin method to find a band insulator with staggered density, a spin-density-wave and a Mott insulating phase. Both the band insulator and the spin-density wave come in various topological varieties. Finally, we calculate the response function relevant for lattice modulation spectroscopy with cold atomic gases in optical lattices.  
\end{abstract}

\date{\today}

\author{Diana Prychynenko}
\affiliation{Institut f\"ur Theoretische Physik, ETH Zurich, CH-8093 Z\"urich, Switzerland}
\affiliation{Institut f\"ur Theoretische Physik, Freie Universit\"at Berlin, Arnimallee 14, 14195 Berlin, Germany}

\author{Sebastian~D.\ Huber}
\affiliation{Institut f\"ur Theoretische Physik, ETH Zurich, CH-8093 Z\"urich, Switzerland}


\maketitle

\section{Introduction}

The motion of quantum mechanical particles can be associated with interesting topological properties. Beyond the standard example of the quantum Hall effect, \cite{Thouless82,Avron85}  lattice problems with zero net magnetic field attracted considerable recent interest. The honeycomb model with complex next-to-nearest neighbor hopping by Haldane \cite{Haldane88} provided the blueprint for a considerable fraction of the current day literature on topological band structures. \cite{Qi11,Hasan10} Despite its pivotal role in the development of this field, a direct experimental implementation was only recently demonstrated with ultra-cold atoms.\cite{Jotzu14} The interesting topological properties of this model arise from the interplay of two energy scales: the strength of the complex next-to-nearest neighbor hopping $t'e^{i\varphi}$, which breaks time-reversal symmetry if $\varphi \notin \lbrace 0,\pi \rbrace$, and the sub-lattice potential $V$, which breaks inversion symmetry. The natural question that poses itself is how an additional energy scale in the form of interactions enriches the picture. 

Interactions can alter the physics of particles on topological band structures profoundly. There are several possible scenarios of how interactions can induce new phases. First, for partially filled bands interactions might stabilize gapped quantum liquids akin the Laughlin states for the fractional quantum Hall effects.\cite{Tang11,Sun11,Neupert11} Another possibility is that the interplay of $t'$, $V$ and an interaction scale $U$ leads to symmetry broken states, where the quasi-particles above these states inherit the underlying band-topology.\cite{He11} 

In this manuscript we discuss how such symmetry broken states can arise at half filling. We explain how they can be described beyond a simple Hartree-Fock theory using slave-particle techniques. Finally, we calculate response functions relevant to current experiments with cold atoms and show how the topological properties of the band structure are revealed. These questions deserve attention as current experiments implement fully tunable honeycomb lattices\cite{Tarruell12,Jotzu14} where both, the Berry curvature of the bands have been measured\cite{Duca14} and interactions effects have been observed.\cite{Uehlinger13} 

In the following, Sec.~\ref{sec:model}, we introduce the concrete model under investigation. We discuss its possible phases and derive them using both a simple Hartree-Fock (Slater determinant) trial wave function as well as a more sophisticated ${\mathds Z}_{2}$ slave-spin method\cite{Huber09a,Ruegg10} which is able to capture interaction effects beyond the physics of Slater determinants. In Sec.~\ref{sec:response} we derive the response functions relevant to the current experiments with ultra-cold fermions.

\section{Ionic Hubbard model at half filling on the honeycomb lattice}
\label{sec:model}

\begin{figure}[b]
\includegraphics{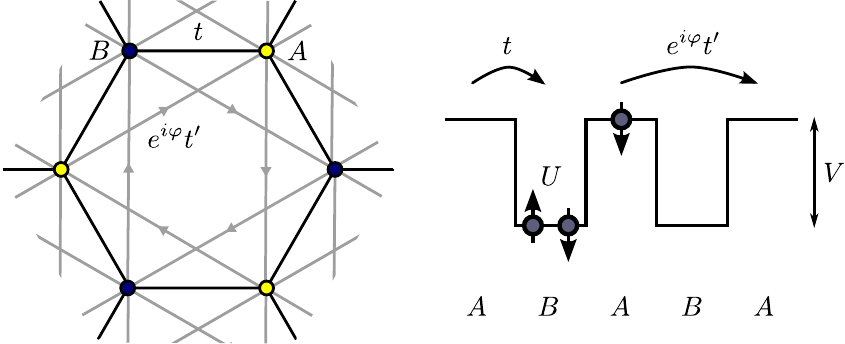}
\caption{
{\bf Setup.} (Left) The honeycomb lattice with its two sub-lattices $A$ and $B$. The gray arrows indicate the phase convention of the next-to-nearest neighbor hopping, see text. (Right) The different terms in the Hamiltonian: the hopping amplitudes $t$ and $t'$, respectively; the sub-lattice potential $V$; and the local repulsion between different spin species $U$.
}
\label{fig:setup}
\end{figure}

\subsection{Model}

We study the ionic Hubbard model on the honeycomb lattice
\begin{multline}
\label{eqn:ionic}
H = 
-\sum_{i,j;\sigma} t_{ij}c_{i\sigma}^{\dag}c_{j\sigma}^{\pdag} 
+
\frac{U}{2} 
\sum_i \bigg(\sum_{\sigma=\uparrow\downarrow}n_{i\sigma}-1\bigg)^2
\\
+\frac{V}{2}\bigg(\!\sum_{i \in A,\sigma} n_{i\sigma}
\!-\!
\sum_{i \in B,\sigma} n_{i\sigma}\!\bigg).
\end{multline}
The operators $c_{i\sigma}^{\dag}$ create fermions in two different spin species $\sigma=\uparrow,\downarrow$ and $t_{ij}$ denote the hoppings on the honeycomb lattice as indicated in Fig.~\ref{fig:setup}. The hopping to the next-to-nearest neighbors is associated with a phase $\phi$, such that the fermion gains $\phi$, when the hopping is performed clockwise around the unit cell. Finally, we have terms proportional to an onsite repulsion $U$ between the different spin-species and a sub-lattice potential $V$. We do not specify a chemical potential as we only consider the case of half filling, i.e., one particle per lattice site where the number of $\uparrow$-fermions equals the number of $\downarrow$-fermions.

Let us discuss the well-known phases of this model. For nearest-neighbor hopping only ($t'=0$) and $V=U=0$, the half-filled system is a semi-metal. The density of states vanishes linearly at the particle-hole symmetric Dirac points at $\mathbf{K}=(2\pi/a)(2/3,0)$ and $\mathbf{K}'=(2\pi/a)(1/3,1/\sqrt{3})$, respectively.\cite{Wallace47} 

Turning on $t'$ breaks the particle-hole symmetry. Moreover, if $\phi\notin \{0,\pm \pi\}$ the system enters a quantum Hall state with Chern numbers in both spin sectors $\mathcal C=(\mathcal C_{\uparrow},\mathcal C_{\downarrow})=\pm(1,1)$.\cite{Haldane88}. An inversion-symmetry-breaking term as the sub-lattice potential, $V\ne 0$, opposes the quantum Hall state and eventually renders the system a simple band insulator with strong density modulation.\cite{Haldane88}

For $V=0$ but $U > U_{\ssm crit}$ the fermions form a spin-density wave (SDW). Note that due to the vanishing density of states at the Dirac point, a finite interaction strength $U_{\ssm crit}$ is need for the SDW to occur.\cite{Sorella92,Honerkamp08} Eventually, for $U\gg t$ the fermions get localized in a Mott insulator and form a Heisenberg anti-ferromagnet.\cite{Martelo96,Raghu08}

How are the transitions between these phases characterized? The onset of the SDW goes along with a symmetry breaking of the spin-rotation symmetry $\mathsf{SU}(2)$ and can be well described within the Ginzburg-Landau framework. The Mott transition on the other hand is only characterized by a qualitative change in the charge fluctuations, concretely by a vanishing charge imbalance between the two sub-lattices. Finally, the transition where the Chern numbers $\mathcal C$ are changing requires necessarily a closing of the excitation gap. We are seeking a method that can capture all these phases and transitions in a unified framework. 

Readers who are not interested in the technical details can skip the next section and directly advance to Sec.~\ref{sec:result}.

\subsection{Method}

In order to describe all aforementioned phases and transitions we employ a slave-spin technique.\cite{Huber09a,Ruegg10} This method, which is tailored to half filling, can track both the excitation spectrum and strongly correlated phases such as the Mott insulator.\cite{orthogonal} \nocite{Nandkishore12} In the following we give a concise account of the slave-spin method and refer the interested reader to Ref.~\onlinecite{Ruegg10} for further details.

The basic building block of the slave-spin method is the introduction of auxiliary degrees of freedom in the form of a constrained slave spin-1/2 (with eigenvalues $I_{i}^{z}=\pm 1/2$) on every site
\begin{equation}
\label{eqn:costraint}
c_{i\sigma} = 2I^x_i f_{i\sigma},\quad I_i^z +\frac{1}{2} = 
\bigg(\sum_{\sigma=\uparrow\downarrow}n_{i\sigma}-1\bigg)^2,
\end{equation}
where $f_{i\sigma}$ are regular fermionic operators and $n_{i\sigma}=f_{i\sigma}^{\dag}f_{i\sigma}^{\pdag}$. The second part of Eq.~(\ref{eqn:costraint}) represents the constraint which slaves the two operators $f_{i\sigma}$ and ${\bf I}_{i}$ to each other. Moreover, it is evident from the constraint that $I_{i}^{z}=1/2$ corresponds to either an {\em empty or a double occupied} site, while $I_{i}^{z}=-1/2$ signals a {\em singly occupied} site. Expressed in the new operators the Hamiltonian reads
\begin{align}
\nonumber
H&=
-\sum_{\langle i,j\rangle,\sigma} 
4t_{ij} I^x_iI^x_jf_{i\sigma}^{\dag}f_{j\sigma}^{\pdag}
+ \frac{U}{2} \sum_{i} I_i^z\\
&\phantom{=}+\frac{V}{2}\bigg(\sum_{i \in A,\sigma} n_{i\sigma}
-
\sum_{i \in B,\sigma} n_{i\sigma}\bigg),
\end{align}
Where we used the constraint to write the interaction part $\propto U$ in the slave-spin sector alone. 

Assuming an ansatz for the ground-state wave-function of the form $|\Psi\rangle=|\Psi_{f}\rangle\otimes |\Psi_{I}\rangle$ we readily obtain the mean-field Hamiltonian
\begin{multline}
H_{\ssm MF} = 
\langle \Psi_{I} | H |\Psi_{I} \rangle +
\langle \Psi_{f} | H |\Psi_{f} \rangle \\
-
\frac{\lambda}{2} \sum_{i} 
\left(I_{z} - 2n_{i\uparrow}n_{i\downarrow} + \sum_{\sigma} n_{i\sigma} \right), 
\end{multline}
where we added a global Lagrange-multiplier $\lambda$ to enforce the the constraint on average. The resulting meanfield Hamiltonians are given by 
\begin{multline}
\label{eqn:slavef}
H_{f} = \sum_{ij} t_{ij}g_{ij} f_{i\sigma}^{\dag}f_{j\sigma} + 
\frac{V}{2}\bigg(\sum_{i \in A,\sigma} n_{i\sigma}
-
\sum_{i \in B,\sigma} n_{i\sigma}\bigg) \\
- \frac{\lambda}{2} \sum_{i\sigma} n_{i\sigma} 
+ \lambda \sum_{i} n_{i\downarrow}n_{i\uparrow}
\end{multline}
and
\begin{equation}
\label{eqn:slaveI}
H_{I} = 
\sum_{ij}t_{ij}\chi_{ij}I_{i}^{x}I_{j}^{x} 
+ \left(\frac{U-\lambda}{2}\right) \sum_{i} I_{i}^{z}
\end{equation}
The two sectors (fermion and slave-sector) are linked via the self-consistency equations for the two renormalization factors 
\begin{align}
\label{eqn:link}
\begin{split}
g_{ij} &=
4\langle \Psi_{I}| I_{i}^{x}I_{j}^{x}|\Psi_{I} \rangle 
\quad \mbox{and} \\
\chi_{ij} &= 
4\sum_{\sigma} 
\left( \langle \Psi_{f}| f_{i\sigma}^{\dag}f_{j\sigma}^{\pdag}|\Psi_{f}\rangle + c.c. \right).
\end{split}
\end{align}

We are now confronted with the problem of solving the two mean-field Hamiltonians (\ref{eqn:slavef}) and (\ref{eqn:slaveI}). To this end we employ a molecular field approximation to the transverse field Ising model (\ref{eqn:slaveI}) and a Hartree-Fock approximation to (\ref{eqn:slavef}). The benefit of using the slave-spin approximation over a direct Hartree-Fock approximation to the original model (\ref{eqn:ionic}) lies in the fact that the slave-spin method allows the interactions to renormalize the hopping strength via $g_{ij}$ and eventually render the system Mott insulating at $g_{ij}=0$.\cite{Brinkman70} 

\subsubsection{Hartree-Fock}

We start with the Hartree-Fock approximation of the fermionic sector. We assume that the ground-state wave-function is a Slater determinant. To parameterize this Slater-determinant we use the parameters of a quadratic Hamiltonian which we determine self-consistently. Our trial Hamiltonian for this purpose can be written as
\begin{equation}
\label{eqn:hf}
H_{\ssm HF} = \sum_{ij} t_{ij}g_{ij} f_{i\sigma}^{\dag}f_{j\sigma} + 
\sum_{i\in A\sigma} \mu_{\sigma}^{A} n_{i\sigma} +  
\sum_{i\in B\sigma} \mu_{\sigma}^{B} n_{i\sigma}.
\end{equation}
This ansatz contains four parameters $\mu_{\sigma}^{\alpha}$ with $\sigma=\uparrow,\downarrow$ and $\alpha=A,B$. The Hamiltonian can be easily diagonalized to find the full spectrum and eigenstates. We then minimize the energy in the ground state $|\Psi_{\ssm HF} \rangle$ of the Hartree-Fock Hamiltonian (\ref{eqn:hf}) 
\begin{equation}
\frac{\partial \langle \Psi_{\ssm HF} | H_{f} | \Psi_{\ssm HF} \rangle}{\partial \mu_{\sigma}^{\alpha}}=0.
\end{equation}
This minimization yields the self-consistency equations
\begin{align}
\label{eqn:sc1}
\mu_{\sigma}^{A} &= 
\phantom{-}V + \frac{\lambda}{2} - 
\lambda  \langle \Psi_{\ssm HF} | n_{i\bar\sigma}^{A}| \Psi_{\ssm HF} \rangle, \\
\label{eqn:sc2}
\mu_{\sigma}^{B} &= 
-V + \frac{\lambda}{2} - 
\lambda  \langle \Psi_{\ssm HF} | n_{i\bar\sigma}^{B}| \Psi_{\ssm HF} \rangle.
\end{align}
Here $\bar\sigma$ denotes the opposite spin to $\sigma$. Moreover, due to the translationally invariant ansatz (\ref{eqn:hf}), the density $ \langle \Psi_{\ssm HF} |n_{i\sigma}^{\alpha}| \Psi_{\ssm HF} \rangle$ depends only on the sub-lattice index $\alpha=A,B$, not on $i$. 

The variational parameters $\mu_{\sigma}^{\alpha}$ are conjugate to the densities $\rho_{\sigma}^{\alpha}=\langle \Psi_{\ssm HF} | n_{i\sigma}^{\alpha}| \Psi_{\ssm HF} \rangle$. As we constrain ourselves to half-filling, $\sum_{\sigma\alpha}\rho_{\sigma}^{\alpha}=2$, and zero net magnetization, $\sum_{\alpha}\rho_{\uparrow}^{\alpha}=\sum_{\alpha}\rho_{\downarrow}^{\alpha}$, only two of them are independent. For convenience we introduce two independent parameters 
\begin{align}
m &= \rho_{\downarrow}^{A}
+\rho_{\uparrow}^{B}-(\rho_{\uparrow}^{A}+\rho_{\downarrow}^{B}), \\
\Delta n & = \rho_{\downarrow}^{A}+\rho_{\uparrow}^{A} - 
(\rho_{\uparrow}^{B}+\rho_{\downarrow}^{B}).
\end{align}
While $m$ characterizes a staggered magnetization and hence a breaking of $\mathsf{SU}(2)$ symmetry, $\Delta n$ describes a staggered density, indicating a charge imbalance between the sublattices as long as $\Delta n \ne 0$. 

For the Hartree-Fock approximation to the original model (\ref{eqn:ionic}), solving the self-consistency equations (\ref{eqn:sc1}) and (\ref{eqn:sc2}) provides us with the mean-field phase diagram.\cite{He11} For the case of the slave-spin method we also need to solve the spin-part and find a solution of both the spin and the fermion sector linked by (\ref{eqn:link}).

\subsubsection{Molecular-field approximation}

To solve the spin-part we employ a molecular field approximation to the transverse field Ising model (\ref{eqn:slaveI}). To this end, we replace $I_{i}^{x}I_{j}^{x}\rightarrow \langle I_{i}^{x}\rangle I_{j}^{x}$, based on the assumption that the fluctuations from the mean value are small. This renders the slave-spin sector essentially a single-site problem in which we have to self-consistently determine $\langle I^{x}\rangle$. The single-site problem reads
\begin{equation}
\label{eqn:mf}
H_{I}^{\ssm MF} = {\bf h} \cdot {\bf I}
\end{equation}
with
\begin{equation}
{\bf h} =
 \left[\underbrace{-(zt\chi+z't'\chi')}_{\bar{\chi}}
 \langle I^{x}\rangle,0,\frac{U-\lambda}{2}\right].
\end{equation}
Here, $z$ and $z'$ are the number of nearest and next-to-nearest neighbors and $\chi$ and $\chi'$ the respective expectation values in the fermionic sector (\ref{eqn:link}). Solving (\ref{eqn:mf}) amounts to a simple rotation in spin space under the constraint that we recover a self-consistent solution for $\langle I^{x}\rangle$ which yields
\begin{equation}
\langle I^{x} \rangle = \sqrt{\frac{1}{4}-\left(\frac{U-\lambda}{2\bar\chi}\right)^{2}}.
\end{equation}
\begin{figure*}[t!]
\includegraphics{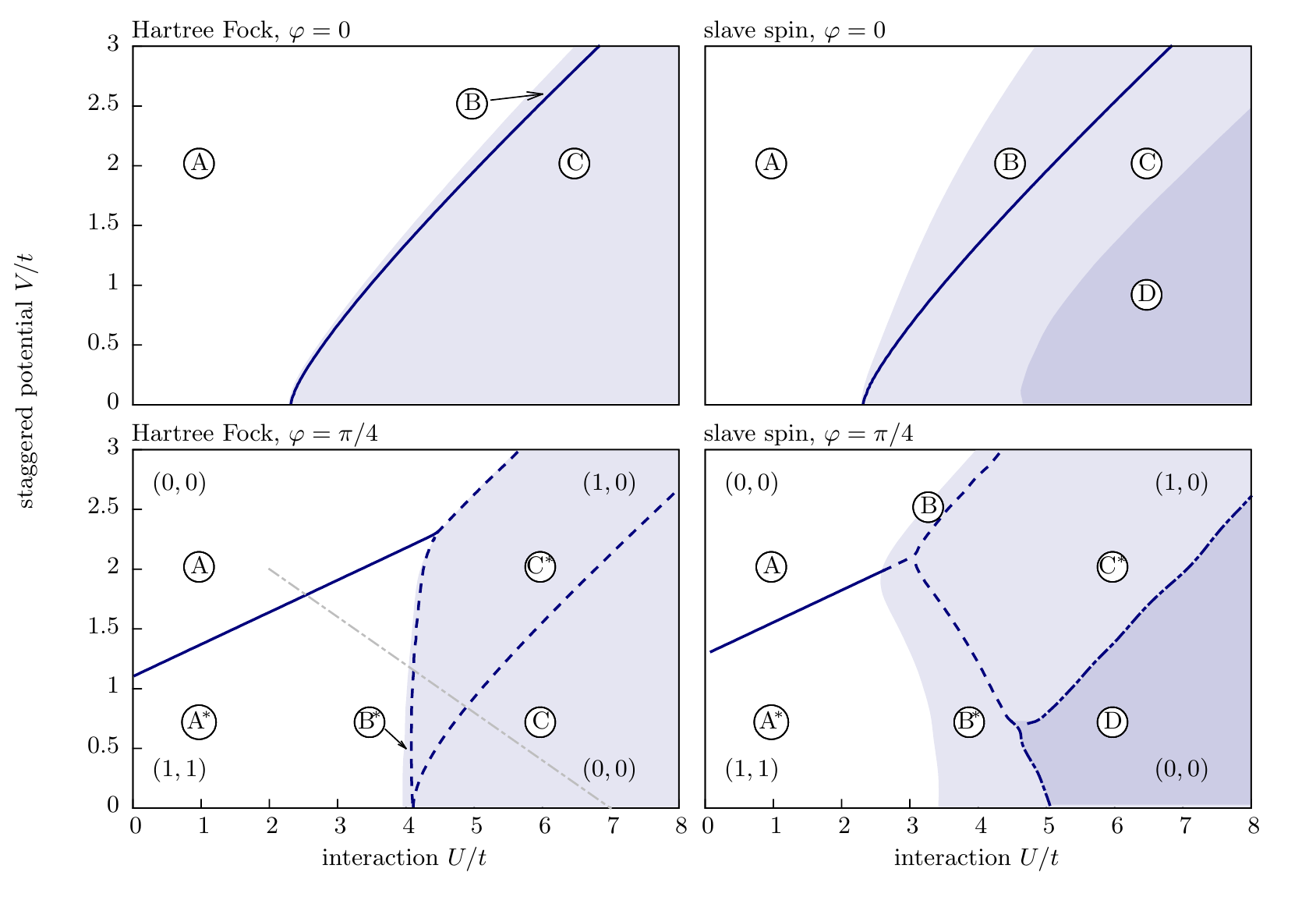}
\caption{
{\bf Phase diagram.} The phase diagram of the topological honeycomb model as derived via a Hartree-Fock approximation (left panels) and via a slave-spin method (right panels) as a function of sub-lattice potential $V$ and interaction strength $U$. The phases $A$ are simple band insulators while $B$ stands for a spin-density wave (SDW) phase which is adiabatically connected to the phase $A$. The phases $C$ are SDW phases that are separated from the band insulator via a gap-closing in the excitation spectrum (solid or dashed lines). The phases $D$ are Mott insulators where strong interaction effects renormalize the hopping to zero. For broken time-reversal symmetry (bottom panels), the different phases are additionally labelled by their respective Chern numbers $\mathcal C=(\mathcal C_{\uparrow},\mathcal C_{\downarrow})$, where all phases with at least one non-zero Chern number  are labelled with a star. See text for details.
}
\label{fig:pd}
\end{figure*}

\subsubsection{Combinig the ${\bf I}$ and $f$ sectors}
 
Let us review our progress so far. First, we introduced slave-spins $\bf I$ and $f$-fermions. We then assumed a product wave function $|\Psi\rangle = |\Psi_{f}\rangle\otimes|\Psi_{I}\rangle$. Solving both sectors individually (via a Hartree-Fock and a molecular-field approximation, respectively) we obtained solutions parametrized by
\begin{align*}
\mbox{$f$-sector:}
\quad&m(g_{ij},\lambda),\;\Delta n(g_{ij},\lambda),\; 
\chi(g_{ij},\lambda),\; \chi'(g_{ij},\lambda),\\
\mbox{$\bf I$-sector:}\quad&\langle I^{x}\rangle(\bar\chi,\lambda).
\end{align*}

The procedure consists in using (\ref{eqn:sc1}) and (\ref{eqn:sc2}) to resolve the self-consistency conditions. Owing to the single-site nature of the slave-spin sector the hopping renormalization factor simplifies to
\begin{equation}
\label{eqn:gij}
g_{ij} =g
= 4 \langle I_{i}^{x} I_{j}^{x}\rangle \approx 4\langle I^{x} \rangle^{2} = 
\left[1-\left(\frac{U-\lambda}{\bar\chi}\right)^{2}\right].
\end{equation}
The remaining issue is to determine $\lambda$. To this end we use the following trick $g=4\langle I^{x}\rangle^{2}=1-4\langle I^{z}\rangle^{2}$.  To further simplify this expression we use the exact constraint, introduced in Eq.~(\ref{eqn:costraint}), linking $I^{z}$ to fermionic properties. After straight-forward algebra we find
\begin{equation}
\label{eqn:U}
U = \lambda \pm \bar \chi \frac{(m+\Delta n)(m-\Delta n)}{4}.
\end{equation}
This equation closes the self-consistency loop. The slave-spin sector is completely absorbed in Eqns. (\ref{eqn:gij}) and (\ref{eqn:U}). The fermionic sector can be solved iteratively:
\begin{enumerate}
\item Choose a $\lambda$, $g$ and $\mu_{\sigma}^{\alpha}$.
\item Find the ground state of Eq. (\ref{eqn:hf}).
\item Update $\mu_{\sigma}^{\alpha}$ via (\ref{eqn:sc1}) and (\ref{eqn:sc2}).
\item Calculate $\bar \chi$ and update $g$ via (\ref{eqn:gij}).
\item Iterate the above steps until convergence is reached.
\item After convergence is reached, determine $U$ via (\ref{eqn:U}).
\end{enumerate}
Note that (\ref{eqn:U}) has a $\pm$ ambiguity. As our method has a variational character, for a given $U$ one can compare the mean-field energies to determine which $\lambda$ to choose.

Before we discuss our result, let us make a few comments on the approximations applied so far. First, the assumption of a product state $|\Psi\rangle = |\Psi_{f}\rangle\otimes|\Psi_{I}\rangle$ can be improved via the inclusion of gauge fluctuations in the ${\mathds Z}_{2}$ gauge freedom.\cite{Ruegg10} If the presented mean-field solutions captures the main features of the phase diagram, such gauge-fluctuations should only renormalize the parameters of the mean-field solutions. Second, the restriction to the form of (\ref{eqn:hf}) excludes more complicated symmetry breaking patterns such as superconductivity or incommensurate charge- or spin-density waves. We do not expect such phases to occur at half-filling.\cite{Honerkamp08} Finally, the single-site approximation in the molecular-field solution of the slave-spin-model can be improved via a Holstein-Primakoff spin-wave theory. However, as shown in Ref.~\onlinecite{Huber09a}, this is only needed very close to the Mott transition, or if one is interested in the high-energy excitation spectrum.\cite{Maldague77}

\subsection{Results}
\label{sec:result}

In Fig.~\ref{fig:pd} we show the resulting phase diagrams obtained via the methods outlined above. The value of $t'=0.3\,t$ is fixed and we choose two different values for the phase: $\phi \in \{0,\pi/4\}$. We compare the direct Hartree-Fock calculation\cite{He11} (left panels) to the slave-spin results (right panels).

Let us start with the time-reversal invariant setup for $\phi=0$. The phases labeled by $A$ are simple band insulators with a density modulation $\Delta n \ne 0$ induced by $V$. The lightly shaded regions $C$ indicate a finite SDW order parameter $m$, whereas the dark region $D$ is a Mott insulator with $\Delta n = 0$ and $g=0$. The onset of $m$ is smooth, i.e., happens in a second order transition. The transition between $C$ and $D$ is of first order, which is a known artifact of the slave-spin mean-field theory.\cite{Ruegg10} 

The solid line marks a gap closing for both spin species at both the $K$ and $K'$ points. The region labeled by $B$ is characterized by $m\neq 0$ (and hence a broken spin-rotation symmetry) but is adiabatically connected to the band insulator (no gap closing). To summarize: It is evident that $V$ opposes the instability towards an SDW. Note, that the slave-spin approach enhances the stability of the phase $B$ with respect to the Hartree-Fock results. In general, it stabilizes the SDW ordering towards larger values of the sublattice potential. Moreover, it predicts a Mott insulator which is beyond the reach of the direct Hartree-Fock calculation.

For the time-reversal broken phase ($\phi=\pi/4$) the slave-spin approach predicts a rich phase diagram. In addition to the presence of a staggered magnetization $m$ we can now also have band structures with a non-vanishing Chern numbers $\mathcal C$. For vanishing $U$, we recover the phase diagram of Haldane:\cite{Haldane88} Below a critical $V$ both spin-species have a Chern number $\mathcal C=(1,1)$. We label this phase $A^{*}$. It is separated from the regular band insulator by a gap-closing at $K$. Note, that in the absence of a $\mathsf{SU}(2)$-symmetry breaking both spin species close the gap at the same time as indicated by the solid line. 

Let us turn to the influence of $U$. As for $\phi=0$, at a critical strength $U_{\ssm c}$ a staggered magnetization appears, giving rise to an SDW. If the system starts out in the trivial region $A$, the SDW phase $B$ is also trivial. Coming from $A^{*}$, on the other hand, in phase $B^{*}$ the quasi-particles still have Chern numbers $\mathcal C=(1,1)$. Increasing $U$ further closes the gap at $K$ (dashed line) for one of the two spin species. We end up in a phase $C^{*}$ where only one of the two spin-components has topologically non-trivial excitations $\mathcal C=(1,0)$. Eventually, for yet stronger interactions the gap at $K'$ (dashed line) of the other spin-component closes and we reach a trivial SDW. In the slave-spin frame-work, this transition is always preempted by the first order transition (dash-dotted line) into the (trivial) Mott insulating state $D$. As in the case of the time-reversal invariant situation, the slave-spin approach seems to enhance the stability of the phases $B$ and $B^{*}$ with respect to the Hartree-Fock calculation.

\begin{figure}[t]
\includegraphics{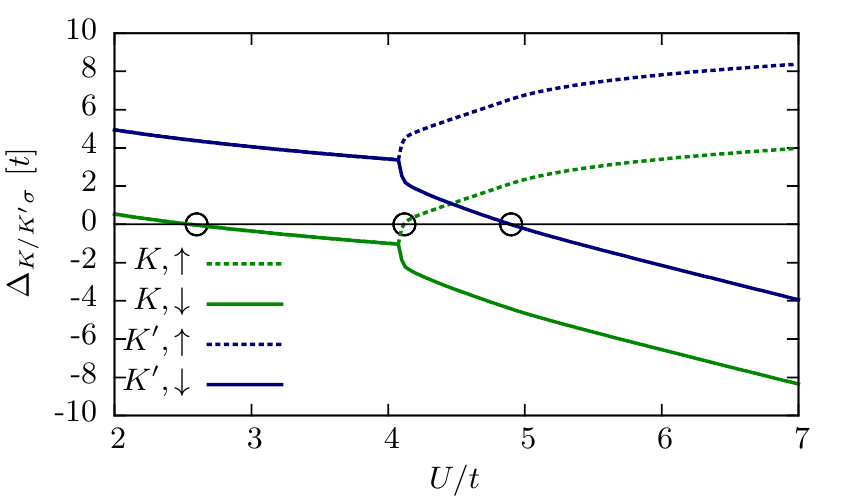}
\caption{
{\bf Gaps.} Evolution of the gaps along dash-dotted line in Fig.~\ref{fig:pd} (bottom left). Green lines denote the gaps at the $K$ point while blue lines are the gaps at the $K'$ points. Solid (dashed) lines indicate the $\uparrow$ ($\downarrow$) gaps, respectively. The circles indicate topological transitions where the Chern number of the respective spin-component changes by one. A deviation of dashed and solid lines indicates the breaking of the spin-rotation symmetry in the spin-density phase.
}
\label{fig:gaps}
\end{figure}

We further discuss the changes in $\mathcal C$. The low-energy Hamiltonian around the Dirac points is given by
\begin{align}
 H_{K\sigma}^{\ssm D} &= \frac{3}{2}t(k_y\tau_x-k_x\tau_y) - 3t^{\prime}\text{cos}(\varphi) - \Delta_{K\sigma} \tau_z, \\
 H_{K'\sigma}^{\ssm D} &= -\frac{3}{2}t(k_y'\tau_x+k_x'\tau_y) - 3t^{\prime}\text{cos}(\varphi) - \Delta_{K'\sigma} \tau_z,
\end{align}
where $k_{x/y}$ and $k_{x/y}'$ denote the deviation from the $K$ and $K'$ points, respectively. We have further defined the gap functions at the two Dirac points
\begin{align}
 \Delta_{K\sigma}&= \frac{U}{2}(\rho_{\bar{\sigma}}^A - \rho_{\bar{\sigma}}^B) - V + 3\sqrt{3}t^{\prime}\text{sin}(\varphi),\\
 \Delta_{K^{\prime}\sigma}&=\frac{U}{2}(\rho_{\bar{\sigma}}^A - \rho_{\bar{\sigma}}^B) - V - 3\sqrt{3}t^{\prime}\text{sin}(\varphi),
\end{align}
within the Hartree-Fock approximation. 
In Fig.~\ref{fig:gaps}, we plot the evolution of $\Delta_{K/K',\sigma}$ along the dash-dotted line in Fig.~\ref{fig:pd}. It is evident that for $m=0$ the Chern numbers of both spin species changes together. Also visible is that the onset of $m\neq0$, where the value of the gap for the two  spin-species starts to deviate, does not coincide with the gap-closing transition. This statement refers to the gap for single-particle fermionic excitations. The breaking of the spin-rotation symmetry certainly involves the appearance of an emergent bosonic Goldstone mode.

In summary, we find that the phase diagram of Ref.~\onlinecite{He11} survives the inclusion of strong interactions via the slave-spin method. Moreover, the renormalization of the hopping $g$ seems to stabilize the SDW phases $B$, $B^{*}$, and $C^{*}$, where both $m \ne 0$ and $\Delta n \ne 0$. 

The qualitative features of the phase diagrams are the same for the two methods. Therefore, we use the simple Hartree-Fock variant to calculate the response functions in the following. The same calculation within the slave-spin framework can be performed but is considerably more involved while yielding qualitatively similar results.

\section{Response functions}
\label{sec:response}

One of the standard probing techniques for strongly interacting cold atoms is the lattice modulation spectroscopy.\cite{Stoferle04, Jordens08, Huber09a, Endres12, Kollath06a} In this probing scheme, the depth of the optical lattice is modulated with a given frequency $\omega$. For the case of fermionic atoms it has been shown that the most sensitive probe is to count the number of doubly occupied sites after ramping up a very deep optical lattice.\cite{Jordens08,Uehlinger13} The relevant response function is given by\cite{Huber09a,Kollath06a}
\begin{equation}
\label{eqn:chi}
\Xi(\omega)  = 
\sum_{m} \langle m | \delta D | m \rangle |\langle m | K | 0 \rangle|^{2} 
\delta(\omega-\omega_{m0}),
\end{equation}
where $\delta D = \sum_{i} n_{i\uparrow}n_{i\downarrow} - \langle 0 | \sum_{i} n_{i\uparrow}n_{i\downarrow} | 0\rangle$ measures the change in double-occupancy with respect to the ground state $|0\rangle$, $K=\sum_{ij\sigma} t_{ij} c_{i\sigma}^{\dag} c_{j\sigma}^{\pdag}$ is the kinetic energy operator, and $\hbar\omega_{m0}$ is the energy difference between the ground and the excited state $|m\rangle$.

\begin{figure}[tb]
\includegraphics{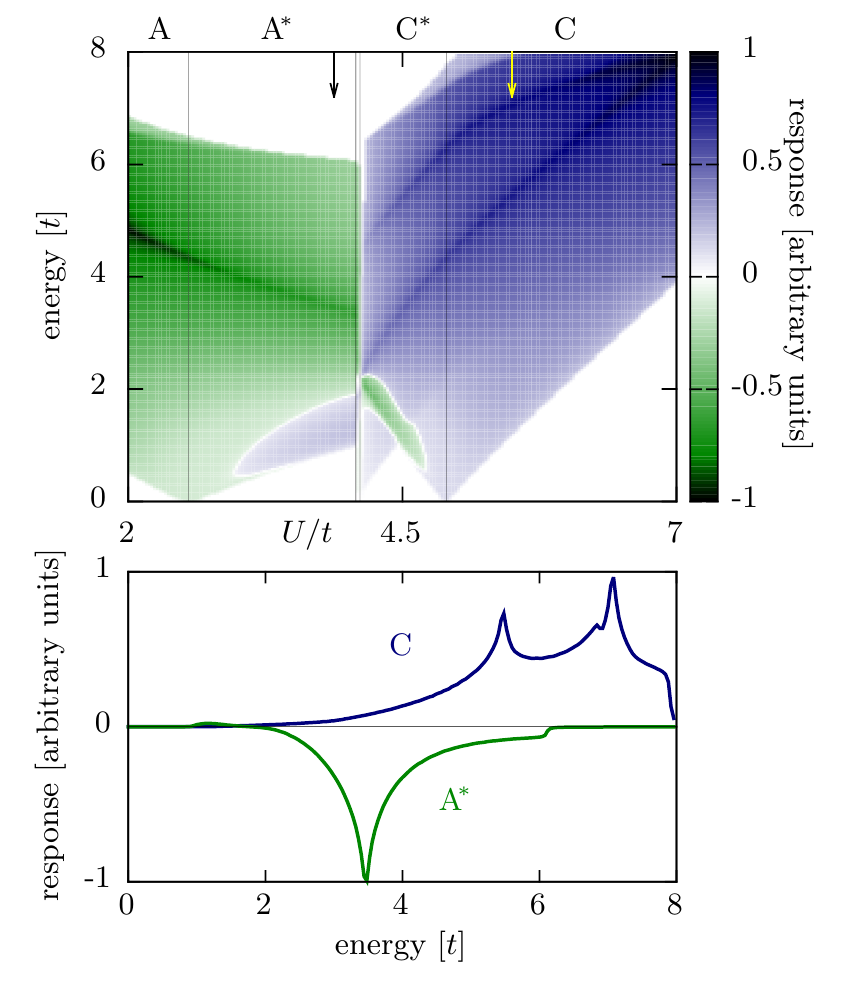}
\caption{
{\bf Response.} (Top panel) Intensity plot of the response function $\Xi(\omega)$ along the dash-dotted line (horizontal axis) in Fig.~\ref{fig:pd} as a function of frequency (vertical axis). Shades of violet indicate a positive signal where the double-occupancy in the excited state is higher than in the ground state. Shades of green quantify a negative signal where the double occupancy is reduced by the excitations. The gap closing transitions between the different phases $A$, $A^{*}$, $C^{*}$, and $C$ are clearly visible. In addition the two topologically non-trivial phases $A^{*}$ and $C^{*}$ {\em can} give rise to an ``inverted band picture'' where, depending on the frequency, both positive and negative signals can be expected. (Lower panel) two cuts through the top panel indicated by the arrows. 
}
\label{fig:response}
\end{figure}

After optimizing the parameters of the Hartree-Fock slater-determinant it is straight forward to evaluate Eq.~(\ref{eqn:chi}). We show the resulting response function along a cut through the phase diagram in Fig~\ref{fig:response}. 

Let us discuss the results. As expected, we find a negative signal in the band-insulator ($A$). The sub-lattice potential $V$ favors doubly occupied sites and the modulation of the lattice induces a depletion of such doublons. In the other extreme, for $U \gg t$ and $U\gg V$, we expect very little double occupancy in the ground state and indeed the modulation leads to a increase in the doublon density, i.e., a positive signal.

The most interesting response is predicted for the topologically non-trivial phases $A^{*}$ and $C^{*}$. First, all gap-closing transitions are clearly visible also in the lattice modulation spectroscopy. Moreover, it is interesting to note that in the two non-trivial phases $A^{*}$ and $C^{*}$, the character of the response changes as a function of energy can change: At low energy (high energy) the response is positive (negative) in some parts of the $A^{*}$ phase. In $C^{*}$ phase this is reversed. Note, however, that this inversion is not in one-to-one correspondence with the phase and hence cannot be used as a strict indication of the topology. However, if observed, such an inversion is a indication of the two phases $A^{*}$ and $C^{*}$.

 \section{Summary and outlook}

We have calculated the phase diagram of the strongly interacting topological honeycomb model using a $\mathds Z_{2}$ slave-spin technique. Our results demonstrate that the simple mean-field diagram of Ref.~\onlinecite{He11} is stable under the inclusion of strong interaction effects. Moreover, we find that all the interesting phases where symmetry breaking in the form of a spin-density-wave and a topological band structure co-exist are enhanced compared to the simple Hartree-Fock results. 

In addition to the ground-state phase diagram we have calculated the response function relevant for recent experiments with cold atoms.\cite{Uehlinger13} Our main finding is that with the lattice-modulation spectroscopy one can see all interesting gap closing transitions responsible for the change in the Chern number $\mathcal C$. Moreover, the topologically non-trivial nature of the ground state {\em can} reveal itself via a change in sign of the response $\Xi(\omega)$ as a function of $\omega$.

\acknowledgments

We acknowledge stimulating discussions with Erez Berg, Evert van Nieuwenburg, Andreas R\"uegg, Roman S\"usstrunk, Murad Tovmasyan, and the lattice team in the group of T. Esslinger. This work was supported in part by National Science Foundation Grant No. PHYS-1066293 and the hospitality of the Aspen Center for Physics. We acknowledge support by the Swiss National Science Foundation.


\begin{thebibliography}{30}
\bibitem[{ort()}]{orthogonal}
\bibinfo{note}{Strictly speaking the original ${\mathds Z}_2$ slave-spin
  method\cite{Ruegg10} does not describe a Mott insulator but rather an
  ``orthogonal metal'', see Ref.~\onlinecite{Nandkishore12}}.
\end{thebibliography}
\end{document}